\documentclass[journal,12pt,onecolumn,draftclsnofoot]{IEEEtran}

\usepackage{bm}
\usepackage{amsmath}
\usepackage{graphicx}
\usepackage{subfigure}
\usepackage{amsfonts}
\usepackage{amssymb}
\usepackage{epstopdf}
\usepackage{algorithm}
\usepackage{algpseudocode}
\usepackage{epsfig}
\usepackage{cite}
\usepackage{threeparttable}
\usepackage{color}
\newtheorem{remark}{Remark}
\DeclareMathOperator*{\argmax}{argmax}

\ifCLASSINFOpdf

\else

\fi

\begin{document}

\title{Sparse Kronecker-Product Coding for Unsourced Multiple Access}

\author{Zeyu Han,~\IEEEmembership{Student Member,~IEEE,}
        Xiaojun Yuan,~\IEEEmembership{Senior Member,~IEEE,}
        Chongbin Xu,~\IEEEmembership{Member,~IEEE,}
        Shuchao Jiang,~\IEEEmembership{Student Member,~IEEE,} \\
        and Xin Wang,~\IEEEmembership{Senior Member,~IEEE}% <-this % stops a space
\thanks{Zeyu Han, Chongbin Xu, Shuchao Jiang, and Xin Wang are with the Key Laboratory for Information Science of Electromagnetic Waves (MoE), Department of Communication Science and Engineering, Fudan University, Shanghai 200433, China (e-mail: \{19210720077, chbinxu, 17110720042, xwang11\}@fudan.edu.cn).}% <-this % stops a space
\thanks{Xiaojun Yuan is with the National Key Laboratory of Science and Technology on Communications, University of Electronic Science and Technology of China,
Chengdu 610000, China (e-mail: xjyuan@uestc.edu.cn).}% <-this % stops a space
}

\markboth{}
{Han \MakeLowercase{\textit{et al.}}: Sparse Kronecker-Product Coding for Unsourced Multiple Access}

\maketitle

\begin{abstract}
In this paper, a sparse Kronecker-product (SKP) coding scheme is proposed for unsourced multiple access. Specifically, the data of each active user is encoded as the Kronecker product of two component codewords with one being sparse and the other being forward-error-correction (FEC) coded. At the receiver, an iterative decoding algorithm is developed, consisting of matrix factorization for the decomposition of the Kronecker product and soft-in soft-out decoding for the component sparse code and the FEC code. The cyclic redundancy check (CRC) aided interference cancellation technique is further incorporated for performance improvement. Numerical results show that the proposed scheme outperforms the state-of-the-art counterparts, and approaches the random coding bound within a gap of only 0.1 dB at the code length of 30000 when the number of active users is less than 75, and the error rate can be made very small even if the number of active users is relatively large.
\end{abstract}


\IEEEpeerreviewmaketitle

\clearpage

\section{Introduction}

\IEEEPARstart{M}{assive} machine-type communication (mMTC) is one of the most important scenarios for future wireless communications \cite{beyond,UMA}. Unsourced multiple access (UMA), first proposed by Polyanskiy in \cite{Unsourced}, provides a new paradigm for mMTC. Different from traditional multiple access protocols, UMA assumes that the access point (AP) only needs to recover the set of messages transmitted by active users. By further assuming that all users adopt the same codebook, the function of the AP reduces to determine the list of active messages in this common codebook. This provides an attractive solution for mMTC. The achievability bound in \cite{Unsourced} was derived based on random coding at the transmitters and maximum likelihood (ML) decoding at the AP. Based on this analysis, it is shown that traditional methods such as ALOHA are very inefficient. Significant performance gains have been reported by some recent designs, like the T-fold ALOHA scheme \cite{foldALOHA} and the low density parity check (LDPC) coding scheme with successive interference cancellation (SIC) \cite{SIC}. Nevertheless, these schemes perform far away from the bound.

Recently, new approaches have been developed to significantly reduce the gap towards the bound in \cite{Unsourced}. For example, the authors in \cite{Tree} proposed to segment the long data packet into several short sections with different sections combined by the tree code, and the best result so far with no power allocation when the number of the active users $K_a\geq250$ is obtained in \cite{AMPTree2}. Combining the ideas of approximate message passing (AMP), tree code, as well as power allocation, further performance improvement is demonstrated by Fengler et al. in \cite{AMPTree1plus}, providing the state-of-the-art results for $K_a\geq225$. An alternative approach is to segment the data packet into a short \emph{head} and a long \emph{body} with the head also serving as (or determining) the identification of the body such as signature (or interleaver) \cite{IDMA, IRSAPolar}. Using polar codes aided by cyclic redundancy check (CRC) and SIC, \cite{Polar2} provides the state-of-the-art solution for $K_a\leq200$.

However, the above two approaches have their own limitations. First, the error correction capability of the segmented tree code is generally weak, and it may cause a substantial performance loss when $K_a$ is small. Second, the complexity of decoding head information is proportional to the size of the random codebook used for head encoding; hence, the codebook cannot be large, which in turn worsens its collision performance when $K_a$ is large.

In this paper, we propose a novel sparse Kronecker-product (SKP) coding scheme. Specifically, the data of each active user is encoded as the Kronecker product of two component codewords, with one component sparse code to facilitate the compressed sensing based multi-user detection and the other component forward-error-correction (FEC) code to obtain coding gain with low complexity. Note that our SKP coding reduces to sparse code multiple access (SCMA) \cite{SCMA} when the component sparse code of SKP does not carry information (and thus can be treated as a sparse spreading sequence). On the other extreme, SKP coding reduces to sparse modulation when the component FEC code does not carry information. Indeed, the extra coding gain of our SKP potentially comes from the enlarged design space of manipulating the information and code-length allocation between the two component codes. At the receiver, an iterative decoding algorithm is developed, consisting of bilinear generalized approximate message passing (BiG-AMP) \cite{BiG1,SSL} based matrix factorization for the decomposition of the Kronecker product and soft-in soft-out decoding for the component sparse code and the FEC code. To suppress the effects of bad initializations in BiG-AMP, the cyclic redundancy check (CRC) aided interference cancellation technique is also incorporated for further performance improvement. Numerical results under the popular setting in \cite{Unsourced} show that the proposed scheme outperforms all the existing schemes, and approaches the random coding bound within a gap of only 0.1 dB at the code length of 30000 when the number of active users is less than 75, and the error rate can be made very low by adjusting the signal-to-noise ratio (SNR) even if the number of active users is relatively large.

The rest of this paper is organized as follows. Section II outlines the system model. Section III delineates the proposed scheme. Numerical results are provided in Section IV. Finally, Section V concludes the paper.

\emph{Notation}: $\mathcal{CN}(\mu,\sigma^2)$ denotes the complex Gaussian distribution with mean $\mu$ and variance $\sigma^2$; ${{\mathcal{CN}}}(a; \mu,\sigma^2)$ denotes the probability density value of random variable $x\sim \mathcal{CN}(\mu,\sigma^2)$ at $x=a$; $\lfloor x \rfloor$ denotes the floor function of $x$; $\otimes$ denotes the Kronecker product; $\mathbb{C}$ denotes the complex number field; and $||\cdot ||_F$ denotes the Frobenius norm.

\section{System Model}

For UMA, we follow the common system model in \cite{Unsourced, foldALOHA, SIC, AMPTree2, AMPTree1plus, IDMA, IRSAPolar, Polar2}. Specifically, consider a Gaussian random access system with $K$ users. Denote by $K_a$ the number of active users in a frame. Each active user transmits a packet of $B$ bits to the AP on the complex channel over $N_{tot}$ real degrees of freedom (rdof) totally, i.e., $N_{tot}/2$ complex channel usages. The received signal $\bm{y}$ at the AP is modeled as

\begin{equation}
\bm{y}=\sum_{k=1}^{K}u_k\cdot \bm{v}_k(\bm{b}_k)+\bm{w} \label{eori}
%\vspace{-0.25cm}
\end{equation}
where $u_k=1$ if the user $k$ is active and $u_k=0$ otherwise, yielding $\sum_{k=1}^{K}u_k=K_a$. Vectors $\bm{b}_k\in\{0,1\}^{B}$, $\bm{v}_k$, and $\bm{w}\sim\mathcal{CN}(0, N_0 \bm{I}_{N_{tot}/2})$ denote the original data packet of user $k$, the transmit signal of user $k$, and the additive white Gaussian noise (AWGN) respectively. All $\{ u_k \}$ and $\{ \bm{b}_k \}$ are assumed to be independent and identically distributed (i.i.d.) with respect to the user index $k$. All active users $\left\{k\left|\right.u_k=1\right\}$ are subject to the power constraint $||\bm{v}_k||^2\leq P$. The average bit SNR is defined as

\begin{equation}
\frac{E_b}{N_0}\triangleq\frac{P}{B N_0}. \label{eebn0}
%\vspace{-0.2cm}
\end{equation}

In the UMA framework, the task of the receiver is to decode all packets based on $\bm{y}$ and provide the resulting list $\mathcal{L}(\bm{y})$, which includes at most $K_a$ packets. The per-user probability of error (PUPE) is defined as \cite{Unsourced}

\begin{equation}
P_e\triangleq\frac{1}{K_a}\sum_{j=1}^{K_a}\Pr\left\{\mathbb{E}_j\right\} \label{eper}
\end{equation}
where $j$ denotes the index of active users' list; i.e., $u_j=1, \forall j$. $\mathbb{E}_j\triangleq\left\{\bm{b}_j\notin \mathcal{L}(\bm{y})\right\}\cup\left\{\exists i\neq j,\ \bm{b}_j=\bm{b}_i\right\}$ is the error event of active user $j$.

Given $N_{tot}$ and $B$, the design target is to optimize the coding schemes to meet the performance requirement $P_e\leq \varepsilon$ with the lowest ${E_b}/{N_0}$ value.

%\vspace{-0.4cm}
\section{Sparse Kronecker-Product Coding}

In this section, we develop the proposed SKP coding scheme. We start with the encoding process in Sec. III-A, introduce the decoding operations in Sec. III-B, and then propose the CRC-aided interference cancellation technique for further performance improvement in Sec. III-C.

\subsection{Encoder}
\begin{figure}[ht]
\centering
\includegraphics[width=0.7\textwidth]{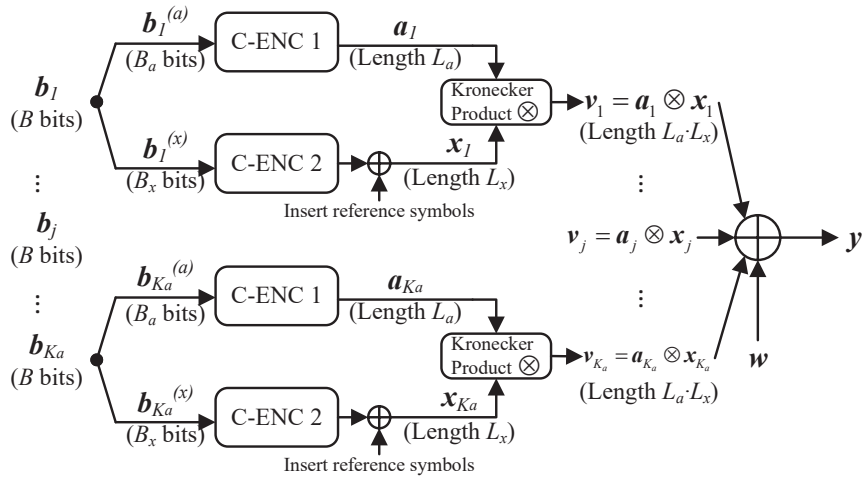}\\ %{extra/encoder.pdf}\\
\caption{The encoding process. ``C-ENC'' stands for the component encoder.}
\label{encoder}
\end{figure}

The encoding process of the proposed design is shown in Fig. \ref{encoder}. For any user $j$, the $B$ bits in one packet are divided into two parts: the first $B_a$-bit $\bm{b}_j^{(a)}$ is encoded (including channel coding and modulation) as a length-$L_a$ vector and the remaining $(B_x=B-B_a)$-bit $\bm{b}_j^{(x)}$ is encoded as a length-$(L_x=\lfloor N_{tot}/(2L_a) \rfloor)$ vector. The transmit signal is obtained by the Kronecker product of the two vectors. Then the received signal $\bm{y}$ at the AP can be written as

\begin{equation}
\bm{y} = \sum_{j=1}^{K_a}\bm{a}_j(\bm{b}_j^{(a)}) \otimes \bm{x}_j(\bm{b}_j^{(x)})+\bm{w} \label{eenc}
\end{equation}
where $\bm{a}_j$ denotes the encoded vector with message $\bm{b}_j^{(a)}$, and $\bm{x}_j$ denotes the other encoded vector with message $\bm{b}_j^{(x)}$.

Let $\bm{A} = [\bm{a}_1, \cdots, \bm{a}_{K_a}]\in \mathbb{C}^{L_a\times K_a}$, $\bm{X} = {[\bm{x}_1, \cdots, \bm{x}_{K_a}]}^T\in \mathbb{C}^{K_a\times L_x}$, and reshape $\bm{y}$ to $\bm{Y}\in \mathbb{C}^{L_a\times L_x}$, and $\bm{w}$ to $\bm{W}\in \mathbb{C}^{L_a\times L_x}$ accordingly. The received signal at the AP can be rewritten in a matrix form as

\begin{equation} \label{eencmat}
\bm{Y} = \sum_{j=1}^{K_a}\bm{a}_j \bm{x}_j^T + \bm{W} = \bm{A}\bm{X} + \bm{W}.
%\vspace{-0.2cm}
\end{equation}

The structures of the matrices $\bm{A}$ and $\bm{X}$ in (\ref{eencmat}) are determined by the selection of the two component codes. At the receiver, matrix factorization can be used to estimate $\bm{A}$ and $\bm{X}$ (and thereby the original packets $\{ \bm{b}_j \}$) from $\bm{Y}$. Notice that sparse structure is usually preferred for such factorization. For this reason, we require that $\bm{A}$ is sparse. Further considering the complexity of utilizing such structured information in the decoding algorithm, we construct $\bm{A}$ by index modulation (IM) \cite{IdxM} and $\bm{X}$ by FEC codes, as detailed next.

For each column $\bm{a}_j$ in the matrix $\bm{A}$, the length-$L_a$ vector is divided into $g_a$ segments with each segment having length $(I_a = L_a / g_a)$ and only one nonzero element taking its value from the constellation $\mathcal{S}$. It can be verified that each segment (i.e., IM symbol) carries $(\log_{2}{I_a}+\log_{2}{|\mathcal{S}|})$ bits and $\bm{a}_j$ carries $B_a=g_a(\log_{2}{I_a}+\log_{2}{|\mathcal{S}|})$ bits with sparsity ratio $\lambda = 1/{I_a}$. Notice that each $\bm{a}_j$ consists of multiple IM symbols. The probability that different active users select the same $\bm{a}_j$ can be very low with the coding parameters properly selected.

For each row $\bm{x}_j^T$ in matrix $\bm{X}$, the FEC code is utilized. As the packet in mMTC is usually short, the FEC code having good performance at short code-length should be selected. To facilitate the iterative decoding, the selected code should also have a soft decoding algorithm. To this end, a tail-biting convolutional code (CC), used in LTE and LTE-A, is adopted. Additionally, to eliminate the phase ambiguity of matrix factorization, $e_{Re\!f}$ reference symbols (denoted by $s_p$) are included in each $\bm{x}_j$. The tail-biting CC is then used to encode the $B_x$ bits, and the encoded sequence is modulated with the constellation $\mathcal{S}$ to generate a length-$(L_x\!-e_{Re\!f})$ vector.

Finally, the signal of each user is generated by the Kronecker product $\bm{a}_j\otimes \bm{x}_j$, and is transmitted over the channel.

\subsection{Decoder}
Theoretically, the maximum \textit{a posteriori} (MAP) decoding can be used to recover $\{\bm{b}_j\}_{j=1}^{K_a}$ by solving the following optimization problem:

\begin{equation}
\begin{split}
%\{ \hat{\bm{b}}_j \} = \mathop{\argmax}_{ \{ \bm{b}_j \}} & \ {\exp\left(-\frac{||\bm{Y}-\bm{A(\{\bm{b}_j\})X(\{\bm{b}_j\})}||_F^2}{N_0}\right)} \\
\{ \hat{\bm{b}}_j \} = \mathop{\argmax}_{ \{ \bm{b}_j \}} & \ {\exp\left(-\frac{||\bm{Y}-\bm{A}(\{\bm{b}_j^{(a)}\})\bm{X}(\{\bm{b}_j^{(x)}\})||_F^2}{N_0}\right)} \cdot p_A \left(\bm{A}(\{\bm{b}_j^{(a)}\}) \right)\cdot p_X \left(\bm{X}(\{\bm{b}_j^{(x)}\}) \right)\\
\end{split}
\label{emap}
\end{equation}
where $p_A(\bm{A})$ and $p_X(\bm{X})$ denote the \textit{a priori} distributions of $\bm{A}$ and $\bm{X}$ respectively, which are determined by their specific coding structures.

\begin{figure*}[ht]
\centering
\includegraphics[width=0.98\textwidth]{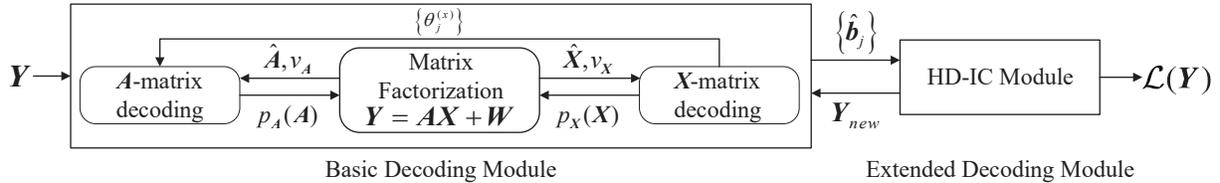}\\ %{extra/decoder.pdf}\\
\caption{Structure of the decoder.}
\label{decoder}
\end{figure*}

Iterative decoding provides a low complexity alternative \cite{BiG1,SSL,Rayleigh,softSIC}. In particular, for the bilinear problem in (\ref{emap}), the algorithms in \cite{BiG1,SSL} can be applied when $\bm{A}$ and $\bm{X}$ are uncoded. In our design, both $\bm{A}$ and $\bm{X}$ have their specific coding structures, and accurate messages are hard to obtain. To solve the problem, we propose the iterative decoding scheme illustrated in Fig. \ref{decoder}. The basic decoding module contains three submodules: a matrix factorization submodule to decompose $\bm{A}$ and $\bm{X}$ from $\bm{Y}$ by ignoring the coding constraints of $\bm{A}$ and $\bm{X}$, an IM decoder to refine the estimate of $\bm{A}$, and an FEC decoder to refine the estimate of $\bm{X}$.

\subsubsection{Matrix factorization submodule}
This submodule estimates $\bm{A}$ and $\bm{X}$ based on the received signal $\bm{Y}$ and the feedback information of $\bm{A}$ and $\bm{X}$ from the other two submodules.

The BiG-AMP algorithm is adopted for the matrix factorization, which is an extension of generalized AMP (GAMP) to the bilinear system model. Similar to GAMP, BiG-AMP is derived according to the message passing principle and Gaussian message approximation based on the Taylor series expansion and the central-limit theorem with arbitrary input distributions.

BiG-AMP consists of four main steps. Denote by $\bm{Z}=\bm{A}\bm{X}$. Firstly, the messages of $\bm{Z}$ are calculated based on the messages of $\bm{A}$ and $\bm{X}$. Secondly, by further combining the observation $\bm{Y}$, the \textit{a posteriori} means and variances of the elements in $\bm{Z}$ are obtained. Thirdly, the messages of $\bm{A}$ (or $\bm{X}$) are calculated by using the messages of $\bm{Z}$ and $\bm{X}$ (or $\bm{A}$) according to the constraint $\bm{Z}=\bm{A}\bm{X}$. Finally, the \textit{a posteriori} means and variances of the elements in $\bm{A}$ (or $\bm{X}$) are obtained by further combining the \textit{a priori} information of $\bm{A}$ (or $\bm{X}$). The four steps iterate to obtain the estimates of $\bm{A}$ and $\bm{X}$.

More details about the BiG-AMP algorithm are shown in the Appendix A.

As shown in Fig. \ref{decoder}, the feedback messages $p_A(\bm{A}), p_X(\bm{X})$ are given as the inputs of BiG-AMP. Note that the input messages are element-wise, and $a_{l_a,j}\in \{0\} \bigcup \mathcal{S}$ and $x_{l_x,j}\in \mathcal{S}$. The element-wise marginal distributions of $\{a_{l_a,j}\}$ and $\{x_{l_x,j}\}$ are given in the following forms

\begin{equation}
p_{a_{l_a,j}}(a_{l_a,j}) = \Phi_{a_{l_a,j}}(0)\delta(a_{l_a,j}) + \sum_{i=1}^{|\mathcal{S}|}\Phi_{a_{l_a,j}}(s_i)\delta(a_{l_a,j}-s_i) \label{eap1}
%\vspace{-0.2cm}
\end{equation}

\begin{equation}
p_{x_{l_x,j}}(x_{l_x,j}) = \sum_{i=1}^{|\mathcal{S}|} \Phi_{x_{l_x,j}}(s_i)\delta(x_{l_x,j}-s_i) \label{eap2}
%\vspace{-0.2cm}
\end{equation}
where $s_i \in \mathcal{S}$, and $\Phi_{a}(s_i)$ denotes the probability $\Pr {\left\{a=s_i\right\}}$. In the first iteration, there is no feedback from the other submodules and the initial values in (\ref{eap1}) and (\ref{eap2}) are set as $\Phi_{a_{l_a,j}}(0) = 1\!-\!\lambda$, $\Phi_{a_{l_a,j}}(s_i) = {\lambda}/{|\mathcal{S}|}$, and $\Phi_{x_{l_x,j}}(s_i) = {1}/{|\mathcal{S}|}$.

The outputs of BiG-AMP to other submodules are the extrinsic element-wise Gaussian messages of $\bm{A}$ and $\bm{X}$, characterized by mean matrix $\hat{\bm{A}}$ and variance matrix $v_{\bm{A}}$, as well as mean matrix $\hat{\bm{X}}$ and variance matrix $v_{\bm{X}}$, in an element-wise form as

\begin{equation} \label{eapinit}
\left\{
\begin{split}
a_{l_a,j} & \sim \mathcal{CN}\left(\hat{a}_{l_a,j}, v_{a_{l_a,j}}\right) \\
x_{l_x,j} & \sim \mathcal{CN}\left(\hat{x}_{l_x,j}, v_{x_{l_x,j}}\right).
\end{split}
\right.
\end{equation}

The detailed calculation of (\ref{eapinit}) is given by Eqs. (A9)--(A12) in Table II in the Appendix A.

With $\{ \hat{\bm{A}}, v_{\bm{A}}, \hat{\bm{X}}, v_{\bm{X}} \}$ in (\ref{eapinit}) and the structured information of $\bm{A}$ and $\bm{X}$ in Sec. III-A, the estimates of $\bm{A}$ and $\bm{X}$ can be further refined as detailed below.

\subsubsection{$\bm{X}$-matrix decoding submodule}
This submodule refines the estimates of $\bm{X}$ based on the feedback $\{ \hat{\bm{X}}, v_{\bm{X}} \}$ from the matrix factorization submodule and the structured information of $\bm{X}$ described in Sec. III-A. We apply the message passing algorithm \cite{BP} as follows.

Before decoding, the phase ambiguity in the estimation is first eliminated based on the reference symbols $s_p$ by finding:

\begin{equation}
\begin{split}
\theta^{(x)}_j & = \mathop{\argmax}_{\theta\in\Theta}\prod_{l'_x=1}^{e_{Re\!f}} {{\mathcal{CN}}}(s_p; \hat{x}_{l'_x,j}e^{i\theta}, v_{x_{l'_x,j}})
%&\frac{1}{\pi{v_{x_{l'_x,j}}}}\exp\left(-\frac{|\hat{x}_{l'_x,j}e^{i\theta}-s_p|^2} {\pi{v_{x_{l'_x,j}}}}\right)\\
%\hat{x}'_{l_x,j} & = \hat{x}_{l_x,j} \cdot e^{i\theta^{(x)}_j}
\end{split}
\label{ecc11}
\end{equation}
where $\Theta$ contains all possible phase ambiguities in $\mathcal{S}$. Then we obtain $\hat{x}'_{l_x,j} = \hat{x}_{l_x,j} \cdot e^{i\theta^{(x)}_j}$ based on $\theta^{(x)}_j$.

Since all the $e_{Re\!f}$ reference symbols take the same value\footnote{Here we calculate the soft information of $\{ x_{l_x,j}\}_{l_x=1}^{e_{Re\!f}}$, rather than using its deterministic value $s_p$, to facilitate the design of the iterative algorithm.}, the messages of $\{ x_{l_x,j}\}_{l_x=1}^{e_{Re\!f}}$ are calculated as:

\begin{equation}
\Phi_{x_{l_x,j}}(s_i) \propto%\frac{1}{|\mathcal{S}|}
\prod_{l'_x=1, l'_x\neq l_x}^{e_{Re\!f}} {{\mathcal{CN}}}(s_i; \hat{x}'_{l'_x,j}, v_{x'_{l'_x,j}}), \forall s_i \in \mathcal{S}.
\label{ecc12}
%\vspace{-0.2cm}
\end{equation}

Finally, the Bahl-Cocke-Jelinek-Raviv (BCJR) algorithm \cite{BCJR} is used for the decoding of the remained $\{\hat{x}'_{l_x,j}\}_{l_x=e_{Re\!f}+1}^{L_x}$. The extrinsic information $\Phi_{x_{l_x,j}}(s_i)$ for each symbol of CC in $\bm{X}$ can be calculated based on the log-likelihood ratio (LLR) of each encoded bit output by the BCJR algorithm.

With $\left\{\Phi_{x_{l_x,j}}(s_i)\right\}$, the distribution $p_X(\bm{X})$ in (\ref{eap2}) is updated to provide a finer estimate of $\bm{X}$.

\subsubsection{$\bm{A}$-matrix decoding submodule}
This submodule refines the estimates of $\bm{A}$ based on the feedback $\{ \hat{\bm{A}}, v_{\bm{A}} \}$ from the matrix factorization submodule, the structured information of $\bm{A}$ described in Sec. III-A, and $\{\theta^{(x)}_j\}$ given by the $\bm{X}$-matrix decoding submodule.

Recall that each column $\bm{a}_j$ in $\bm{A}$ is IM modulated. Denote by $a_{l_a,j}$ the $l_a$-th symbol of user $j$, and $I_f(l_a), I_l(l_a)$ the first and the last indices of the IM segment including index $l_a$, i.e., $[a_{I_f(l_a),j}, \cdots, a_{l_a,j}, \cdots, a_{I_l(l_a),j} ]$ forms an IM symbol. The message of each $a_{l_a,j}$ can be calculated as:

\begin{equation}
\begin{split}
\hat{a}'_{l_a,j} & = \hat{a}_{l_a,j} \cdot e^{- i{\theta^{(x)}_j} } \\
\Phi_{a_{l_a,j}}(\tilde{s}_i) & \propto \sum_{ \bm{d} \in D^{(\tilde{s}_i)}} \prod_{l'_a=I_f(l_a), l'_a\neq l_a}^{I_l(l_a)} {{\mathcal{CN}}}(d_{l'_a}; \hat{a}'_{l'_a,j}, v_{a'_{l'_a,j}})
\end{split}
\label{ess2}
\end{equation}
where $\tilde{s}_i \in \{0\} \bigcup \mathcal{S}$ and $D^{(\tilde{s}_i)}$ denotes the set of IM symbols $\{ [d_{I_f(l_a)}, \cdots, d_{l_a}, \cdots, d_{I_l(l_a)} ] \}$ with $d_{l_a}=\tilde{s}_i$.

With $\left\{\Phi_{a_{l_a,j}}(\tilde{s}_i)\right\}$, the distribution $p_A(\bm{A})$ in (\ref{eap1}) is updated to provide a finer estimate of $\bm{A}$.

The operations of the above three submodules iterate until convergence. With the obtained $\{ \hat{\bm{A}}, v_{\bm{A}}, \hat{\bm{X}}, v_{\bm{X}} \}$, the \textit{a posteriori} distribution of $\left\{\bm{b}_j\right\}$ can be calculated to yield the estimates $\big\{\hat{\bm{b}}_j\big\}$ of $\left\{\bm{b}_j\right\}$.

\subsection{Further Improvement}

Due to the suboptimality of BiG-AMP for matrix decomposition, the aforementioned basic decoding module may suffer from the bad initializations in BiG-AMP. To improve the decoding performance, we further consider the following CRC aided interference cancellation scheme.

Specifically, at the encoder, an $e_{C\!RC}$-bit CRC code is added to each $\bm{x}_j$ at the end of $B_x$ bits, then the $(B_x\!+e_{C\!RC})$ bits are encoded and modulated to generate a length-$(L_x\!-e_{Re\!f})$ vector. At the decoder, the hard decision and interference cancellation (HD-IC) module is then added, as shown in Fig. \ref{decoder}.

In HD-IC module, we first calculate the uncertainty of each data packet. The uncertainty $H_j$ is defined as the entropy of $\hat{\bm{b} }_j= \big[ \hat{\bm{b}}_j^{(a)}; \hat{\bm{b}}_j^{(x)}\big]$ (the soft estimate of $\bm{b}_j=\big[\bm{b}_j^{(a)};\bm{b}_j^{(x)}\big]$). Specifically, for $\bm{b}_j^{(a)}$, denote by $\bm{a}_j$ the encoded codeword of $\bm{b}_j^{(a)}$ which consists of $g_a$ IM symbols, written as $\bm{a}_j = [\bm{a}_{1,j}, \cdots, \bm{a}_{g,j}, \cdots, \bm{a}_{g_a,j}]$, and denote by $\mathbb{D}$ the constellation for each IM symbol. The entropy of $\hat{\bm{b}}_j^{(a)}$ can be readily obtained as

\begin{equation}
%\left\{
\begin{split}
H_j^{(a)} & = \sum_{\bm{b}_{j}^{(a)}\in\{0,1\}^{L_a}}{H(\Pr\{\bm{b}_{j}^{(a)} \big| \hat{\bm{b}}_{j}^{(a)}\})} \\
& = \sum_{\bm{a}_{j}\in \mathbb{D}^{g_a}}{H(\Pr\{\bm{a}_{j} \big| \hat{\bm{a}}_{j}\})} \\
& = \sum_{g=1}^{g_a}\sum_{\bm{a}_{g,j}\in \mathbb{D}}{H(\Pr\{\bm{a}_{g,j} \big| \hat{\bm{a}}_{j}\})}
\end{split}
%\right.
%\quad \forall 1\leq j\leq K_a
\end{equation}
where $H(x) = -x\log_2{x}$.

For $\bm{b}_j^{(x)}$, the BCJR algorithm is used for calculating the LLR of each bit, based on which the entropy of $\hat{\bm{b}}_j^{(x)}$ can be obtained as

\begin{equation}
\left\{
\begin{split}
&\Pr\{b_{l_x,j}^{(x)}=0\} = \frac{\exp(LLR_{l_x,j}^{(x)})}{\exp(LLR_{l_x,j}^{(x)})+1} \\
&H_j^{(x)} = \sum_{l_x=1}^{L_x}\big(H(\Pr\{b_{l_x,j}^{(x)}=0\})+H(1-\Pr\{b_{l_x,j}^{(x)}=0\})\big).
\end{split}
\right.
%\quad \forall 1\leq j\leq K_a
\end{equation}

Finally, the uncertainty $H_j$ can be obtained as

\begin{equation}
H_j = H_j^{(a)} + H_j^{(x)}.
\end{equation}

Then, we make a hard-decision on the packets with relatively high reliability $\big\{\hat{\bm{b}}_j\left|\right.H_j < H_{thr}\big\}$, and check the results by the CRC code. Packets passing CRC (or passing after their most unreliable bits are inverted) are added to pending queue $\mathcal{Q}(\bm{Y})$. If any $\bm{b}_j$ is added to $\mathcal{Q}(\bm{Y})$ within $T_{thr}$ iterations, it is added to the final resulting list $\mathcal{L}(\bm{Y})$, and cancelled from $\bm{Y}$ to get $\bm{Y}_{new}$:

\begin{equation}
\bm{Y}_{new} = \bm{Y} - \bm{a}_j(\bm{b}^{(a)}_j) \cdot \bm{x}_j^T(\bm{b}^{(x)}_j).
\label{sdsic1}
\end{equation}

Finally, the decoder returns to the basic decoding module with $\bm{Y} = \bm{Y}_{new}$, and runs until $|\mathcal{L}(\bm{Y})| = K_a$, or no improvement over $T_{max}$ iterations.

\subsection{Outline of the Scheme}

The overall SKP encoding and decoding algorithms are summarized in Algorithms \ref{oencode} and \ref{odecode} respectively.

\begin{algorithm}[H]
\begin{algorithmic}[1]
\Require
Original data packet $\bm{b}_j$ of active user $j$.%B_a, L_a, I_a, g_a, B_x, L_x, e_{Re\!f}, e_{C\!RC}$ and the generator polynomial of CRC and CC.
\Ensure
The transmit signal $\bm{v}_j$ of active user $j$.
\State Split $\bm{b}_j$ to get the $B_a$-bit $\bm{b}^{(a)}_j$ and $B_x$-bit $\bm{b}^{(x)}_j$.
\State Get $e_{C\!RC}$-bit CRC code for $\bm{b}_j$.
\State Encode $\bm{b}^{(a)}_j$ as a length-$L_a$ vector $\bm{a}_j$ by IM.
\State Encode $\bm{b}^{(x)}_j$ and CRC code by tail-biting CC and modulate the obtained codeword to a length-$(L_x-e_{Re\!f})$ vector $\bm{x}^{(CC)}_j$.
\State Add $e_{Re\!f}$ reference symbols in front of $\bm{x}^{(CC)}_j$ to get $\bm{x}_j$.
\State Get $\bm{v}_j$ by Kronecker product $\bm{a}_j\otimes \bm{x}_j$.
\end{algorithmic}
\caption{Overall SKP encoding algorithm}
\label{oencode}
\end{algorithm}

The complexity of the encoding algorithm is $\mathcal{O}(B)$, which is very low and especially suitable for mMTC scenarios.

\begin{algorithm}[H]
\begin{algorithmic}[1]
\Require
$\bm{Y}, N_0, K_a$, and all encoding parameters.%$\mathcal{S}$, and $\lambda$.
\Ensure
Resulting list $\mathcal{L}(\bm{Y})$ ($|\mathcal{L}(\bm{Y})| \leq K_a$).
\State Initial $\mathcal{Q}(\bm{Y})=\varnothing$; $\mathcal{L}(\bm{Y})=\varnothing$; $T_{iter}=0$.
\Repeat
\State Run basic decoding module to obtain the estimates $\{ \hat{\bm{A}}, v_{\bm{A}}, \hat{\bm{X}}, v_{\bm{X}} \}$ using (\ref{eap1})--(\ref{ess2}).
\State Do soft decision to obtain $\big\{\hat{\bm{b}}_j\big\}$.
\State Run HD-IC module based on $\big\{\hat{\bm{b}}_j\big\}$ to update $\mathcal{Q}(\bm{Y})$, $\mathcal{L}(\bm{Y})$, and get $\bm{Y}_{new}$ using (\ref{sdsic1}).
\State Let $\bm{Y} = \bm{Y}_{new}, T_{iter}=0$ if $\bm{Y}_{new} \neq \bm{Y}$, otherwise $T_{iter}=T_{iter}+1$.
\Until{$|\mathcal{L}(\bm{Y})|=K_a$ or $T_{iter}>T_{max}$}.
\end{algorithmic}
\caption{Overall SKP decoding algorithm}
\label{odecode}
\end{algorithm}

The complexity of the decoding algorithm, which is dominated by that with the matrix factorization, is $\mathcal{O}(L_aL_x)$ (normalized for each active user). This complexity is generally low since $L_aL_x \le N_{tot}/2$. Hence the sizes of the codebooks used for $\bm{A}$ and $\bm{X}$ can be both selected flexibly to improve the performance.

\begin{remark}
For SKP coding, each component codeword can be seen as the spreading signature of the other. The key difference here is that a spreading signature does not carry information in a conventional coding scheme, but here both components of the SKP code carry information. Particularly, when the information carried by component $\bm{a}_j$ reduces to zero (i.e., $\bm{a}_j$ is fixed), the SKP code reduces to the well-known SCMA scheme \cite{SCMA}; on the other extreme, when the information carried by component $\bm{x}_j$ reduces to zero (i.e., $\bm{x}_j$ is fixed), the SKP code reduces to an IM scheme \cite{IdxM}. As such, the coding gain of SKP over the conventional SCMA and IM schemes can be achieved by judiciously manipulating the information allocation between the two code components as well as the corresponding code lengths.
\end{remark}

\section{Numerical Results}

Here we use the popular setting in \cite{Unsourced} to compare performance of the SKP coding scheme with existing state-of-the-art schemes \cite{AMPTree2, AMPTree1plus, IRSAPolar, Polar2}. Assume $25\leq K_a \leq 300$, $N_{tot} = 30000$ (i.e., $L_aL_x\leq 15000$), $B = 100$, and $\varepsilon = 0.05$ ($N_{tot} = 26229, B = 89$ in \cite{AMPTree1plus}). The parameters of the encoder are set as in Table \ref{setting}, where GP stands for the generator polynomial of tail-biting CC ($R=1/2$) in octal form, and $I_a$ (i.e. $\lambda = 1/{I_a}$) is chosen empirically to make a good balance between the coding gain of the single user and the multiple access capability of the whole system. Additionally, we use $\pi/4$-quadrature phase shift keying (QPSK) modulation, i.e., $\mathcal{S} = \{\pm \frac{\sqrt{2}}{2}\pm \frac{\sqrt{2}}{2}i\}$, and set $H_{thr} = 32$, $T_{thr} = 4$ and $T_{max} = 15$ in the decoder.

\begin{table}[H]
\centering
\caption{parameters of the encoder}
\label{setting}
\resizebox{0.67\textwidth}{!}{
\begin{threeparttable}
	\begin{tabular}{|c||c|c|c|c|c|c|c|c|c|}
		\hline
        $K_a$ & $\lfloor B_a\rfloor$\tnote{1} & $L_a$ & $g_a$ & $I_a$ & $B_x$ & $L_x$ & GP & $e_{Re\!f}$ & $e_{C\!RC}$\\
        \hline
        25 -- 125  & 37 & 220 & 5 & 44 & 63 & 68 & [561 753] & 4 & 1\\
        \hline
        150        & 30 & 192 & 4 & 48 & 70 & 78 & [561 753] & 5 & 3\\
        \hline
        175 -- 200 & 23 & 174 & 3 & 58 & 77 & 86 & [561 753] & 5 & 4\\
        \hline
        225 -- 250 & 16 & 160 & 2 & 80 & 84 & 93 & [133 171] & 4 & 5\\
		\hline
        275 -- 300 & 16 & 160 & 2 & 80 & 84 & 93 & [23 33]   & 4 & 5\\
        \hline
	\end{tabular}
\begin{tablenotes}
\item[1] We set $\lfloor B_a\rfloor\leq B_a < \lfloor B_a\rfloor+1$ ($B_a=g_a(\log_{2}{I_a}+\log_{2}{|\mathcal{S}|})$) to match the setting and compare with other schemes fairly.
\end{tablenotes}
\end{threeparttable}}
\end{table}

\begin{figure}[H]
\centering
\includegraphics[width=0.7\textwidth]{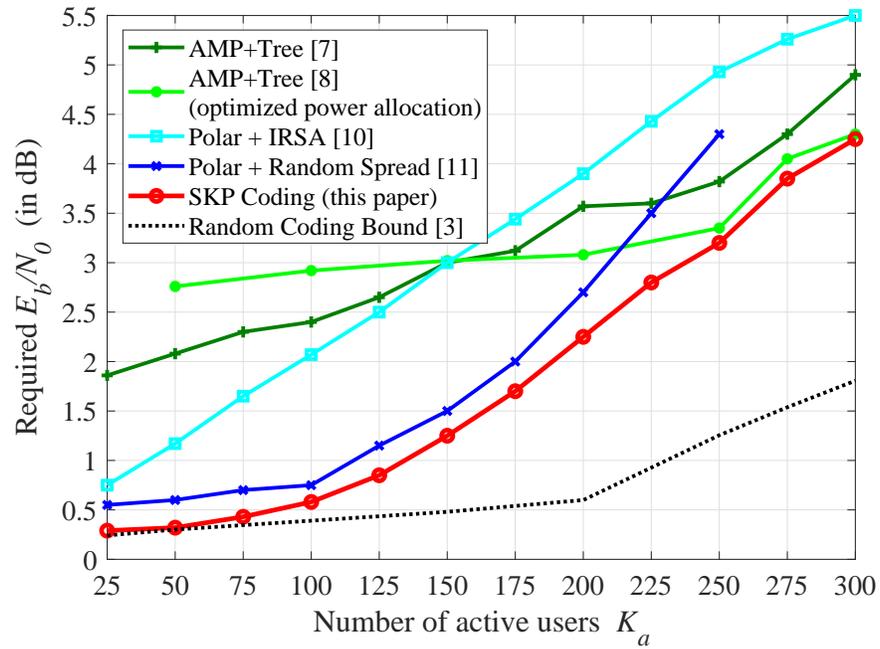}
\caption{Performance comparison of different schemes.}
\label{results}
\end{figure}

The required SNR with the proposed SKP coding scheme for the given $K_a$ is shown in Fig. \ref{results}. It is observed that our scheme outperforms all existing schemes in the full range of $25\leq K_a\leq 300$, and narrows the gap towards the random coding bound to less than 0.1 dB when $K_a\leq 75$.

\begin{figure}[H]
\centering
\includegraphics[width=0.7\textwidth]{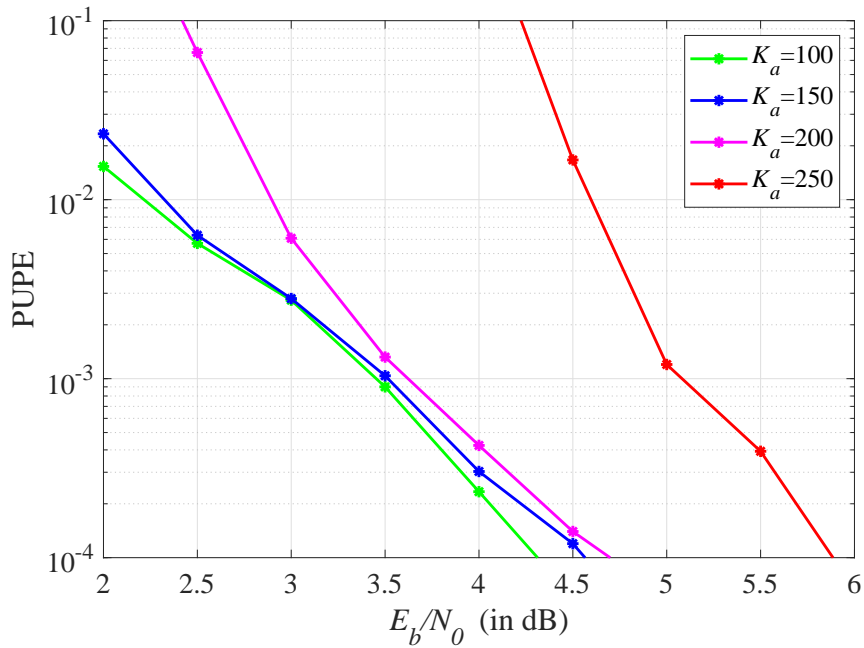}
\caption{The PUPE performance under different $E_b/N_0$ values. $\lfloor B_a\rfloor=30$, $L_a=192$, $g_a=4$, $I_a=48$, $B_x=70$, $L_x=78$, GP = [23 33], $e_{Re\!f}=4$ and $e_{C\!RC}=4$; $H_{thr}=16$, $T_{thr}=2$ and $T_{max}=3$.}
\label{PUPE}
\end{figure}

Additionally, different from the limited-size codebook used for the short heads in \cite{IRSAPolar, Polar2}, the size of the codebook used for $\bm{A}$ and $\bm{X}$ in the SKP coding scheme can be very large under appropriate parameters; as a result, it provides a potentially very low per-user probability of error (PUPE) even when $K_a$ is large, as in Fig. \ref{PUPE}.

\section{Concluding Remarks}
In this paper, we develop a novel SKP coding scheme for unsourced multiple access. The data of each active user is encoded as the Kronecker-product of one sparse IM code and one conventional channel code. At the receiver, an iterative decoding algorithm consisting of matrix factorization for the decomposition of the Kronecker product and two individual decoders for the two component codes is derived, and the CRC aided interference cancellation technique is further adopted for performance improvement. Numerical results show that the proposed scheme outperforms the existing schemes, and approaches the random coding bound within 0.1 dB when the number of active users is less than 75.

Looking forward, there are a number of possible extensions of the work in this paper. For example, in this paper, we assumed a single-antenna AWGN channel between the AP and each user. How to extend the proposed design to multiple-antenna fading channels will be an interesting research direction to pursue in our future work.

\ifCLASSOPTIONcaptionsoff
  \newpage
\fi

\appendices
\section{Details of the BiG-AMP algorithm}
The Bilinear Generalized Approximate Message Passing (BiG-AMP) algorithm \cite{BiG1} is to estimate $\bm{A}$ and $\bm{X}$ based on the noisy observation of $\bm{Z}=\bm{A}\bm{X}$ denoted by $\bm{Y}$, the conditional distribution $p_{\textsf{y}_{ml}|\textsf{z}_{ml}}(y_{ml}|z_{ml})$, and the \textit{a priori} distribution of $\bm{A}$ and $\bm{X}$. For the special AWGN case of $\bm{Y}=\bm{Z}+\bm{W}$, the whole BiG-AMP algorithm is shown in TABLE \ref{e1}. The derivation of the BiG-AMP algorithm is based on sum-product message passing.

The whole algorithm can be divided into four parts. Firstly, in (A1)--(A4), the likelihood distribution of each $z_{ml}$ is obtained, denoted by the Gaussian approximation $\mathcal{CN}(\hat{p}_{ml},\nu^p_{ml})$. Secondly, in (A5)--(A6), the \textit{a posteriori} mean and variance of each $z_{ml}$ are obtained. Thirdly, in (A7)--(A12), the likelihood distributions of $x_{nl}$ and $a_{mn}$ are obtained, denoted by the Gaussian approximations $\mathcal{CN}(\hat{x}_{nl},\nu^x_{nl})$ and $\mathcal{CN}(\hat{a}_{mn},\nu^a_{mn})$ respectively. Finally, in (A13)--(A16), the \textit{a posteriori} mean and variance of each $x_{nl}$ and $a_{mn}$ are obtained, denoted by $\hat{r}_{nl},\nu^r_{nl}$ and $\hat{q}_{mn},\nu^q_{mn}$ respectively. In BiG-AMP algorithm, all the distributions are passed based on message passing algorithm, and the above four parts iterate until convergence, and finally the likelihood distribution in the form of the element-wise Gaussian distribution, and the \textit{a posteriori} mean and variance of each $x_{nl}$ and $a_{mn}$ are determined.

Initially, an initial value of $\{\nu^r_{nl}, \hat{r}_{nl}, \nu^q_{mn}, \hat{q}_{mn}\}$ should be chosen. One approach (used in our scheme) is choosing randomly based on the \textit{a priori} distribution of $\bm{A}$ and $\bm{X}$.

\begin{table}[H]
\caption{The BiG-AMP algorithm}
\label{e1}
\begin{equation*}
\begin{array}{|lrcl@{}r|}\hline
  \multicolumn{2}{|l}{\textsf{definitions:}}&&&\\
  &p_{\mathsf{z}_{ml}|\mathsf{p}_{ml}}(z|\hat{p};\nu^p)
   &\triangleq& \frac{p_{\mathsf{y}_{ml}|\mathsf{z}_{ml}}(y_{ml}|z) \,\mathcal{CN}(z;\hat{p},\nu^p)}
	{\int_{z'} p_{\mathsf{y}_{ml}|\mathsf{z}_{ml}}(y_{ml}|z') \,\mathcal{CN}(z';\hat{p},\nu^p)} &\\
  &p_{\mathsf{r}_{nl}|\mathsf{x}_{nl}}(r|\hat{x};\nu^x)
   &\triangleq& \frac{p_{\mathsf{x}_{nl}}\!(r) \,\mathcal{CN}(r;\hat{x},\nu^x)}
        {\int_{r'}p_{\mathsf{x}_{nl}}\!(r') \,\mathcal{CN}(r';\hat{x},\nu^x)}&\\
  &p_{\mathsf{q}_{mn}|\mathsf{a}_{mn}}(q|\hat{a};\nu^a)
    &\triangleq& \frac{p_{\mathsf{a}_{mn}}\!(q) \,\mathcal{CN}(q;\hat{a},\nu^a)}
         {\int_{q'}p_{\mathsf{a}_{mn}}\!(q') \,\mathcal{CN}(q';\hat{a},\nu^a)}&\\
  \multicolumn{2}{|l}{\textsf{initialization:}}&&&\\
  &\forall m,l:
   \hat{s}_{ml}(0) &=& 0 &\\
  &\forall m,n,l: \textsf{choose~} &
  \multicolumn{2}{l}{\nu^r_{nl}(1), \hat{r}_{nl}(1), \nu^q_{mn}(1), \hat{q}_{mn}(1)} &\\
  \multicolumn{2}{|l}{\textsf{for $t=1,\dots T_\textrm{max}$}}&&&\\
  &\forall m,l:
   \bar{\nu}^p_{ml}(t)
   &=& \textstyle \sum_{n=1}^{N} \left[|\hat{q}_{mn}(t)|^2 \nu^r_{nl}(t) + \nu^q_{mn}(t) |\hat{r}_{nl}(t)|^2\right] \!\!& \text{(A1)}\\
  &\forall m,l:
   \bar{p}_{ml}(t)
   &=& \textstyle \sum_{n=1}^{N} \hat{q}_{mn}(t) \hat{r}_{nl}(t) & \text{(A2)}\\
    &\forall m,l:
   \nu^p_{ml}(t)
   &=& \textstyle \bar{\nu}^p_{ml}(t) + \sum_{n=1}^{N} \nu_{mn}^q(t) \nu^r_{nl}(t) & \text{(A3)}\\
  &\forall m,l:
   \hat{p}_{ml}(t) &=&
   \textstyle \bar{p}_{ml}(t) - \hat{s}_{ml}(t\!-\!1)\bar{\nu}^p_{ml}(t)& \text{(A4)}\\
  &\forall m,l:
   \nu^z_{ml}(t) &=&
   \textstyle \textrm{var}\{\mathsf{z}_{ml}|\mathsf{p}_{ml}\!=\!\hat{p}_{ml}(t);\nu^p_{ml}(t)\} & \text{(A5)}\\
  &\forall m,l:
    \hat{z}_{ml}(t) &=&
   \textstyle  \textrm{E}\{\mathsf{z}_{ml}|\mathsf{p}_{ml}\!=\!\hat{p}_{ml}(t);\nu^p_{ml}(t)\} & \text{(A6)}\\
  &\forall m,l:
   \nu^s_{ml}(t) &=&
   \textstyle {(1 -  \nu^z_{ml}(t)/\nu^p_{ml}(t))/\nu^p_{ml}(t)}  & \text{(A7)}\\
  &\forall m,l:
   \hat{s}_{ml}(t) &=&
   	\textstyle ( \hat{z}_{ml}(t) - \hat{p}_{ml}(t))/\nu^p_{ml}(t) & \text{(A8)}\\
 &\forall n,l:
   \nu^x_{nl}(t)
   &=& \textstyle \big(\sum_{m=1}^{M} |\hat{q}_{mn}(t)|^2 \nu^s_{ml}(t)
	\big)^{-1} & \text{(A9)}\\
  &\forall n,l:
   \hat{x}_{nl}(t)
   &=& \textstyle \hat{r}_{nl}(t) ( 1 - \nu^x_{nl}(t) \sum_{m=1}^{M} \nu^q_{mn}(t) \nu^s_{ml}(t)   ) &\\
   &&&\qquad+ \nu^x_{nl}(t) \sum_{m=1}^{M} \hat{q}_{mn}^*(t)
	\hat{s}_{ml}(t)  & \text{(A10)}\\
  &\forall m,n:
   \nu^a_{mn}(t)
   &=& \textstyle \big(\sum_{l=1}^{L} |\hat{r}_{nl}(t)|^2 \nu^s_{ml}(t)
	\big)^{-1} & \text{(A11)}\\
  &\forall m,n:
   \hat{a}_{mn}(t)
   &=& \textstyle \hat{q}_{mn}(t)(1 - \nu^a_{mn}(t) \sum_{l=1}^{L} \nu^r_{nl}(t) \nu^s_{ml}(t)    ) &\\
   &&&\qquad+ \nu^a_{mn}(t) \sum_{l=1}^{L} \hat{r}_{nl}^*(t)
	\hat{s}_{ml}(t)  & \text{(A12)}\\
  &\forall n,l:
	\nu^r_{nl}(t\!+\!1) &=&
		 \textrm{var}\{\mathsf{r}_{nl}|\mathsf{x}_{nl}\!=\!\hat{x}_{nl}(t); \nu^x_{nl}(t)\} & \text{(A13)}\\
  &\forall n,l:
    \hat{r}_{nl}(t\!+\!1) &=&
    	 \textrm{E}\{\mathsf{r}_{nl}| \mathsf{x}_{nl}\!=\!\hat{x}_{nl}(t); \nu^x_{nl}(t)\} & \text{(A14)}\\
  &\forall m,n:
	\nu^q_{mn}(t\!+\!1) &=&
		 \textrm{var}\{\mathsf{q}_{mn}|\mathsf{a}_{mn}\!=\!\hat{a}_{mn}(t); \nu^a_{mn}(t)\}& \text{(A15)}\\
  &\forall m,n:
	\hat{q}_{mn}(t\!+\!1) &=&
		 \textrm{E}\{\mathsf{q}_{mn}|\mathsf{a}_{mn}\!=\!\hat{a}_{mn}(t); \nu^a_{mn}(t)\} & \text{(A16)}\\
   \multicolumn{4}{|c}{\textsf{if $\sum_{m,l} |\bar{p}_{ml}(t) - \bar{p}_{ml}(t\!-\!1)|^2 \le \tau_\textrm{BiG-AMP} \sum_{m,l} |\bar{p}_{ml}(t)|^2$, {\textsf{stop}}}}&\\
    \multicolumn{2}{|l}{\textsf{end}}&&&\\\hline
\end{array}
\end{equation*}
\end{table}


\begin{thebibliography}{99}

\bibitem{beyond}
X. Chen, D. W. K. Ng, W. Yu, E. G. Larsson, N. Al-Dhahir, and R. Schober, ``Massive access for 5G and beyond,'' \emph{IEEE J. Sel. Areas Commun.}, vol. 39, no. 3, pp. 615-637, Mar. 2021.

\bibitem{UMA}
Y. Wu, X. Gao, S. Zhou, W. Yang, Y. Polyanskiy, and G. Caire, ``Massive access for future wireless communication systems,'' \emph{IEEE Wireless Commun.}, vol. 27, no. 4, pp. 148-156, Aug. 2020.

\bibitem{Unsourced}
Y. Polyanskiy, ``A perspective on massive random-access,'' in \emph{Proc. IEEE Int. Symp. Inf. Theory (ISIT)}, Aachen, 2017, pp. 2523-2527.

\bibitem{foldALOHA}
O. Ordentlich and Y. Polyanskiy, ``Low complexity schemes for the random access Gaussian channel,'' in \emph{Proc. IEEE Int. Symp. Inf. Theory (ISIT)}, Aachen, 2017, pp. 2528-2532.

\bibitem{SIC}
A. Vem, K. R. Narayanan, J. Cheng, and J.-F. Chamberland, ``A user-independent serial interference cancellation based coding scheme for the unsourced random access Gaussian channel,'' in \emph{Proc. IEEE Inf. Theory Workshop (ITW)}, Kaohsiung, 2017, pp. 121-125.

\bibitem{Tree}
V. K. Amalladinne, A. Vem, D. K. Soma, K. R. Narayanan, and J.-F. Chamberland, ``A coupled compressive sensing scheme for uncoordinated multiple access.'' [Online]. Available: https://arxiv.org/abs/1809.04745

\bibitem{AMPTree2}
V. Amalladinne, A. Pradhan, C. Rush, J.-F. Chamberland, and K. R. Narayanan, ``Unsourced random access with coded compressed sensing: Integrating AMP and belief propagation.'' [Online]. Available: https://arxiv.org/abs/2010.04364

\bibitem{AMPTree1plus}
A. Fengler, P. Jung, and G. Caire, ``SPARCs for unsourced random access.'' [Online]. Available: https://arxiv.org/abs/1901.06234

\bibitem{IDMA}
A. Pradhan, V. Amalladinne, A. Vem, K. R. Narayanan, and J.-F. Chamberland, ``A joint graph based coding scheme for the unsourced random access Gaussian channel,'' in \emph{Proc. IEEE Global Commun. Conf. (GLOBECOM)}, Waikoloa, HI, USA, 2019, pp. 1-6.

\bibitem{IRSAPolar}
E. Marshakov, G. Balitskiy, K. Andreev, and A. Frolov, ``A polar code based unsourced random access for the Gaussian MAC,'' in \emph{Proc. IEEE 90th Veh. Technol. Conf. (VTC Fall)}, Honolulu, HI, USA, 2019, pp. 1-5.

\bibitem{Polar2}
A. K. Pradhan, V. K. Amalladinne, K. R. Narayanan, and J.-F. Chamberland, ``Polar coding and random spreading for unsourced multiple access,'' in \emph{Proc. IEEE Int. Conf. Commun. (ICC)}, Dublin, Ireland, 2020, pp. 1-6.

\bibitem{SCMA}
H. Nikopour and H. Baligh, ``Sparse code multiple access,''  in \emph{Proc. IEEE 24th Int. Symp. Pers. Indoor Mobile Radio Commun. (PIMRC)}, London, U.K., 2013, pp. 332-336.

\bibitem{BiG1}
J. T. Parker, P. Schniter, and V. Cevher, ``Bilinear generalized approximate message passing---Part I: Derivation,'' \emph{IEEE Trans. Signal Process.}, vol. 62, no. 22, pp. 5839-5853, Nov. 2014.

\bibitem{SSL}
T. Ding, X. Yuan, and S. C. Liew, ``Sparsity learning-based multiuser detection in grant-free massive-device multiple access,'' \emph{IEEE Trans. Wireless Commun.}, vol. 18, no. 7, pp. 3569-3582, Jul. 2019.

\bibitem{IdxM}
E. Basar, ``Index modulation techniques for 5G wireless networks,'' \emph{IEEE Commun. Mag.}, vol. 54, no. 7, pp. 168-175, Jul. 2016.

\bibitem{Rayleigh}
S. S. Kowshik, K. Andreev, A. Frolov, and Y. Polyanskiy, ``Energy efficient coded random access for the wireless uplink,'' \emph{IEEE Trans. Commun.}, vol. 68, no. 8, pp. 4694-4708, Aug. 2020.

\bibitem{softSIC}
M. Kobayashi, J. Boutros, and G. Caire, ``Successive interference cancellation with SISO decoding and EM channel estimation,'' \emph{IEEE J. Sel. Areas Commun.}, vol. 19, no. 8, pp. 1450-1460, Aug. 2001.

\bibitem{BP}
F. R. Kschischang, B. J. Frey, and H.-A. Loeliger, ``Factor graphs and the sum-product algorithm,'' \emph{IEEE Trans. Inf. Theory}, vol. 47, no. 2, pp. 498-519, Feb. 2001.

\bibitem{BCJR}
L. Bahl, J. Cocke, F. Jelinek, and J. Raviv, ``Optimal decoding of linear codes for minimizing symbol error rate,'' \emph{IEEE Trans. Inf. Theory}, vol. 20, no. 2, pp. 284-287, Mar. 1974.

\end{thebibliography}
\end{document}